\documentclass[a4paper,11pt]{article}

\usepackage{jheppub}
\usepackage[T1]{fontenc}
\usepackage{comment}
\usepackage{multirow}
\usepackage{dcolumn}
\usepackage{bm}
\usepackage{setspace}
\usepackage{graphicx}
\includecomment{printrobustnotes}
\includecomment{printallnotes}

\usepackage{dcolumn}
\usepackage{bm}

\newcommand{\gevcc}[1]  {\ensuremath{#1~\mathrm{GeV}/c^{2}}}

\title{Pulse-shape discrimination between electron and nuclear recoils in a NaI(Tl) crystal}

\author[d,1]{H.S.~Lee,\note{Corresponding author.}}
\author[c]{G.~Adhikari,}
\author[c]{P.~Adhikari,}
\author[b]{S.~Choi,}
\author[g]{I.S.~Hahn,}
\author[a]{E.J.~Jeon,}
\author[b]{H.W.~Joo,}
\author[a]{W.G.~Kang,}
\author[a,b]{G.B.~Kim,}
\author[e]{H.J.~Kim,}
\author[a]{H.O.~Kim,}
\author[b]{K.W.~Kim,}
\author[a]{N.Y.~Kim,}
\author[b]{S.K.~Kim,}
\author[a,c]{Y.D.~Kim,}
\author[a,f]{Y.H.~Kim,}
\author[f]{J.H.~Lee,}
\author[a]{M.H.~Lee,}
\author[a]{D.S.~Leonard,}
\author[a]{J.~Li,}
\author[b]{S.Y.~Oh,}
\author[a]{S.L.~Olsen,}
\author[a]{H.K.~Park,}
\author[f]{H.S.~Park,}
\author[a]{K.S.~Park,}
\author[d]{J.H.~Shim,}
\author[a]{and J.H.~So}

\affiliation[a]{Center for Underground Physics, Institute for Basic Science (IBS), Daejon 305-811, Korea}
\affiliation[b]{Department of Physics and Astronomy, Seoul National University, Seoul 151-747, Korea} 
\affiliation[c]{Department of Physics, Sejong University, Seoul 143-747, Korea}
\affiliation[d]{Department of Physics, Ewha Womans University, Seoul 120-750, Korea} 
\affiliation[e]{Department of Physics, Kyungpook National University, Daegu 702-701, Korea}
\affiliation[f]{Korea Research Institute of Standards and Science, Daejon 205-340, Korea}
\affiliation[g]{Department of Science Education, Ewha Womans University, Seoul 120-750, Korea} 

\emailAdd{hyunsulee@ewha.ac.kr}

\abstract{
We report on the response of a high light-output NaI(Tl) crystal to nuclear recoils induced by neutrons from an Am-Be source and compare the results with the response to electron recoils produced by Compton-scattered 662~keV $\gamma$-rays from a $^{137}$Cs source.  The measured pulse-shape discrimination~(PSD) power of the NaI(Tl) crystal is found to be significantly improved because of the high light output of the NaI(Tl) detector.  We quantify the PSD power with a quality factor and estimate the sensitivity to the interaction rate for weakly interacting massive particles (WIMPs) with nucleons, and the result is  compared with the annual modulation amplitude observed by the DAMA/LIBRA experiment.  The sensitivity to spin-independent WIMP-nucleon interactions based on 100~kg$\cdot$year of data from NaI detectors is estimated with simulated experiments, using the standard halo model.
}

\begin{document}
\maketitle
\flushbottom

\section{Introduction}
The existence of non-baryonic cold dark matter has been widely supported by many astronomical observations~\cite{dm1,dm1-1,dm2,wmap1,wmap2,planck}. The theoretically favored dark matter candidates are weakly interacting massive particles~(WIMPs), which are well motivated by supersymmetric models~\cite{susydm,pdg}. In the constrained minimal supersymmetric standard model, the lightest supersymmetric particle is a WIMP candidate~($\chi$) with an expected mass of $M_{\chi} \geq \gevcc{100}$~\cite{hWIMP}. Recently, a number of experimental observations~\cite{DAMA3, CRESST, CoGent, CDMS-low} could  be interpreted as direct detection signatures of WIMP-nucleon interactions. Moreover recent astronomical gamma-ray observations could be interpreted as indirect signatures of WIMP annihilations near the center of the Galaxy~\cite{astro_ldm}. Therefore unambiguous tests of the direct WIMP detection signals are needed.

Among the direct searches for WIMPs, the DAMA/LIBRA experiment is unique in that it has been consistently reporting positive signals for an annual modulation of the nuclear recoil detection rate since 1998~\cite{DAMA0}, most recently with a 9.3$\sigma$ statistical significance~\cite{DAMA3,DAMA1,DAMA2}. This annual modulation is consistent with WIMP scattering off nuclei inside NaI(Tl) detectors. If the annual modulation signal observed by DAMA/LIBRA is interpreted in the context of the standard halo model~\cite{shm,savage,savage1}, the favored regions of WIMP-nucleon cross-section and WIMP-mass parameter space have been excluded by other direct search experiments~\cite{XENON-low,CDEX,LUX,CDMSlite,kims-prl2,kims-prd}.  This has motivated alternative explanations for the DAMA/LIBRA signal~\cite{alt1,alt2,alt3,davis}. 
However, when non-trivial systematic differences in detector responses~\cite{det1,det2,det3,det4} and reasonable modifications to the commonly used astronomical model for the WIMP distribution~\cite{halo1,halo} are considered, it remains possible to reconcile both the DAMA signal and the null observations~\cite{arina,damaconf}. 
Therefore, it is important to devise experiments that search for the DAMA/LIBRA signal using the same technique but with higher sensitivity. To date, no independent, direct experimental verification of the DAMA/LIBRA signal has been performed.

To reproduce the DAMA/LIBRA observation, several experimental groups have started developing ultra-low-background NaI(Tl) crystals~\cite{anais,dm-ice,kims-nai} with target background levels at or below those measured in the DAMA/LIBRA experiment. If this background reduction is successful, sizable experiments using more than 100~kg of crystals will be performed within the next few years. Accumulating a similar amount of data as that acquired with the DAMA/LIBRA experiment can test the DAMA/LIBRA annual modulation observation. However, this will require a long-term stable data taking over a period of at least three years. 

On the other hand, we can extract a WIMP signature based on differences between the scintillation characteristics  of nuclear and electron recoils using a so-called pulse-shape discrimination~(PSD) analysis. This technique has been applied for WIMP searches with CsI(Tl) crystals~\cite{kims-prl2,kims-plb,kims-prl} and NaI(Tl) crystals~\cite{damapsd,naiad}. In the past, the discrimination power of NaI(Tl) crystals has been much poorer than that of CsI(Tl) crystals~\cite{gerbier,csi-psd1,csi-psd2}. However recently developed NaI(Tl) crystals grown by Alpha Spectra Inc. provide higher light yields~\cite{anais,kims-nai}, and this makes it interesting to study their PSD performance and evaluate their sensitivity. 

In this article, we present a new measurement of the PSD capability of a high light-output NaI(Tl) crystal. We measure the response to nuclear recoils induced by neutrons from a 300-mCi Am-Be neutron source and compare the results with the response to electron recoils induced by a $^{137}$Cs $\gamma$ source. The Seoul National University neutron calibration facility~\cite{csi-psd2} is used for this measurement. Using the results from these measurements, we estimate the sensitivity of a NaI(Tl) dark matter search experiment that uses PSD. 
A model-independent comparison between the expected upper limit on the nuclear recoil rate from the new NaI(Tl) experiment and the DAMA annual modulation amplitude promises a better understanding of the underlying scenarios of current dark matter search results. 

 \section{Measurement of the pulse-shape discrimination}
\begin{figure}
\centering
\includegraphics[width=0.6\textwidth]{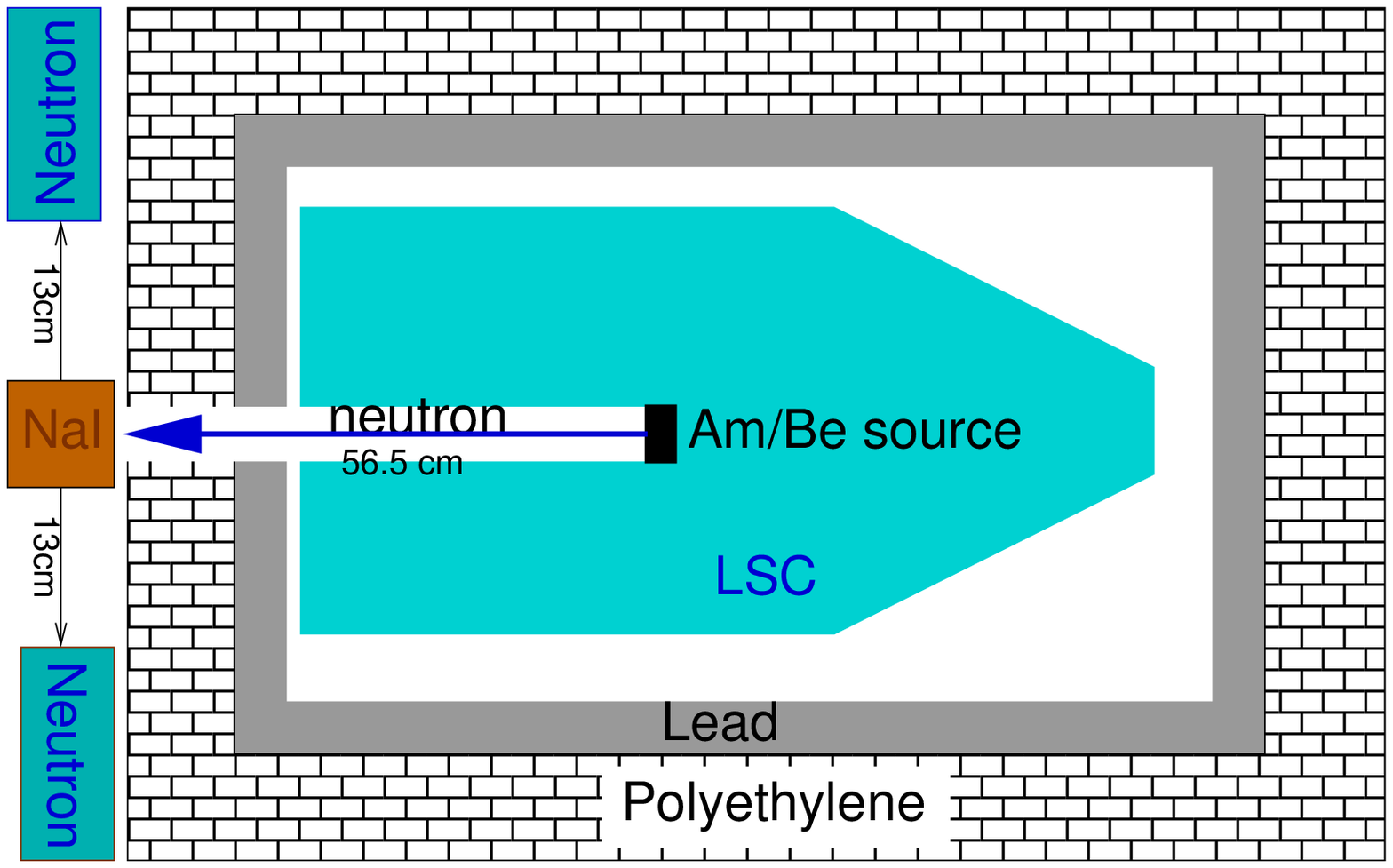}
\caption{A setup for the neutron-nuclei elastic scattering measurements using the Am-Be neutron source.}
\label{fig:ana_nucal_setup}
\end{figure}

 For the neutron-nuclei elastic scattering measurements, the neutron source was surrounded by liquid scintillator~(LSC), lead, and polyethylene shields with a collimated aperture in the crystal direction as shown in Fig.~\ref{fig:ana_nucal_setup}. 
A small 2$\times$2$\times$1.5~cm$^3$ NaI(Tl) crystal made from the same, Alpha Spectra-grown, ingot as a large crystal~(NaI~002) used for background measurements in the Yangyang underground laboratory~\cite{kims-nai} was used as a target. Two 3-inch photomultiplier tubes~(PMTs) that were selected for high quantum efficiency (R12669SEL, Hamamatsu Photonics) were attached to each end of the NaI(Tl) target crystal, which was located at the exit of the collimated aperture of the neutron source. To identify neutrons that elastically scatter in the crystal, two neutron-tagging detectors, consisting of BC501A liquid scintillator contained in cylindrical 0.5~$\ell$ stainless-steel containers, were located at 90$^\circ$ from the line between the neutron source and the NaI(Tl) crystal. The amplified signals from the crystal and the neutron detectors were recorded by 400-MHz flash analog-to-digital converters for 25-$\mu$s time intervals. The trigger condition for the NaI(Tl) crystal was one or more photoelectrons in each PMT within a 200~ns window, which is the same condition that was used for the underground data described in Ref.~\cite{kims-nai}. We additionally required a time coincidence between the NaI(Tl) crystal and one of the neutron detectors to occur within a 2~$\mu$s time window. For the electron recoil measurements, the crystal was irradiated with 662~keV $\gamma$-rays from a $^{137}$Cs source. The data were taken separately with the same setup, except for a lead and polyethylene beam stopper that was placed in the collimated aperture of the neutron source. The trigger condition for the NaI(Tl) crystal was the same; the neutron detectors were not used.

The energy calibration of the NaI(Tl) crystal was performed with 59.54~keV $\gamma$-rays from an $^{241}$Am source. A clustering algorithm described in Ref.s~\cite{kims-plb,csi-psd2} is applied to separate single photoelectrons as individual clusters and reduce the low energy baseline noise. The measured electron-equivalent energy was determined from the charge sum of the identified clusters, and we obtain a light yield for this crystal of approximately 16~photoelectrons/keV, which is consistent with the measured yield for large crystals grown by the same company~\cite{kims-nai,anais}. This light yield is approximately twice larger than that of crystals that were used in previous studies~\cite{gerbier,naiad1,dama0}. 

In the offline analysis, we require a tighter coincidence condition of less than approximately 100~ns between the signals from the NaI(Tl) crystal and the neutron detectors, as discussed in Ref.~\cite{csi-psd2}. Neutron-induced signals in the neutron detectors are identified by taking advantage of the nuclear and electron recoil PSD capabilities of BC501A liquid scintillator~\cite{bc501a,bc501a1}. We recorded neutron data with the Am-Be neutron source over a three-month period.

\begin{figure*}
\begin{center}
\begin{tabular}{cc}
\includegraphics[width=0.45\textwidth]{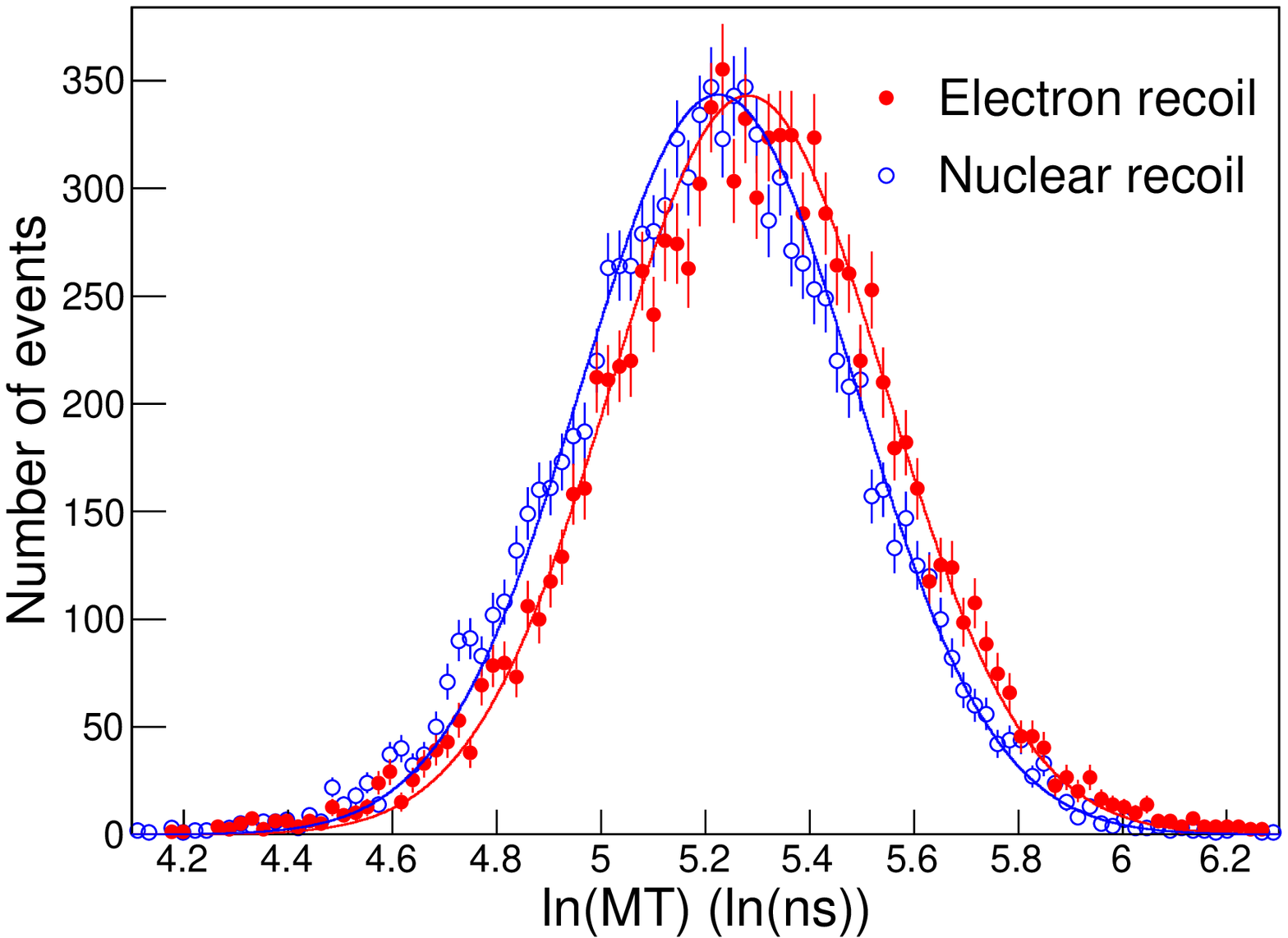}&
\includegraphics[width=0.45\textwidth]{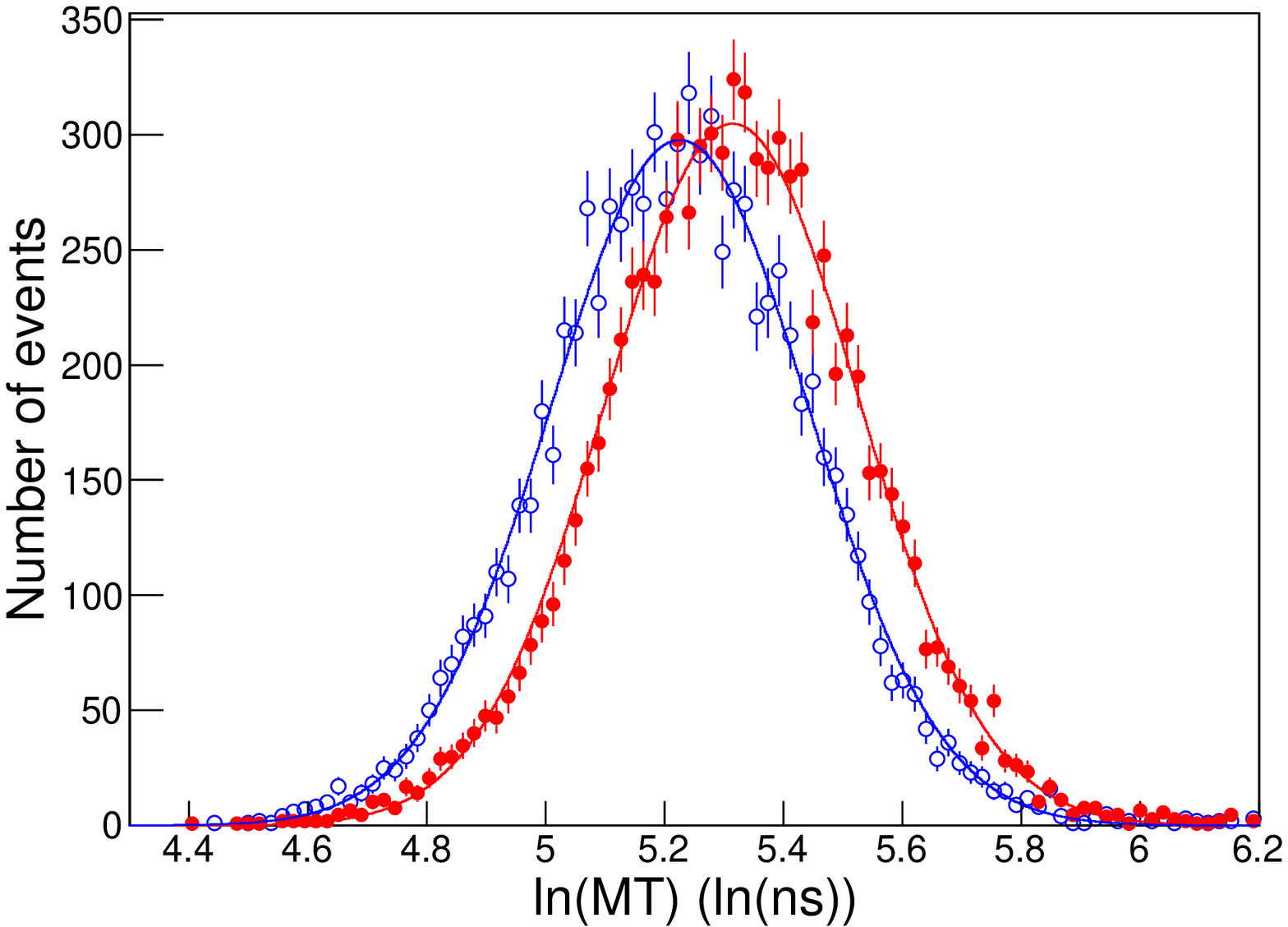}\\
(a) 1-2~keV & (b) 2-3~keV \\
\includegraphics[width=0.45\textwidth]{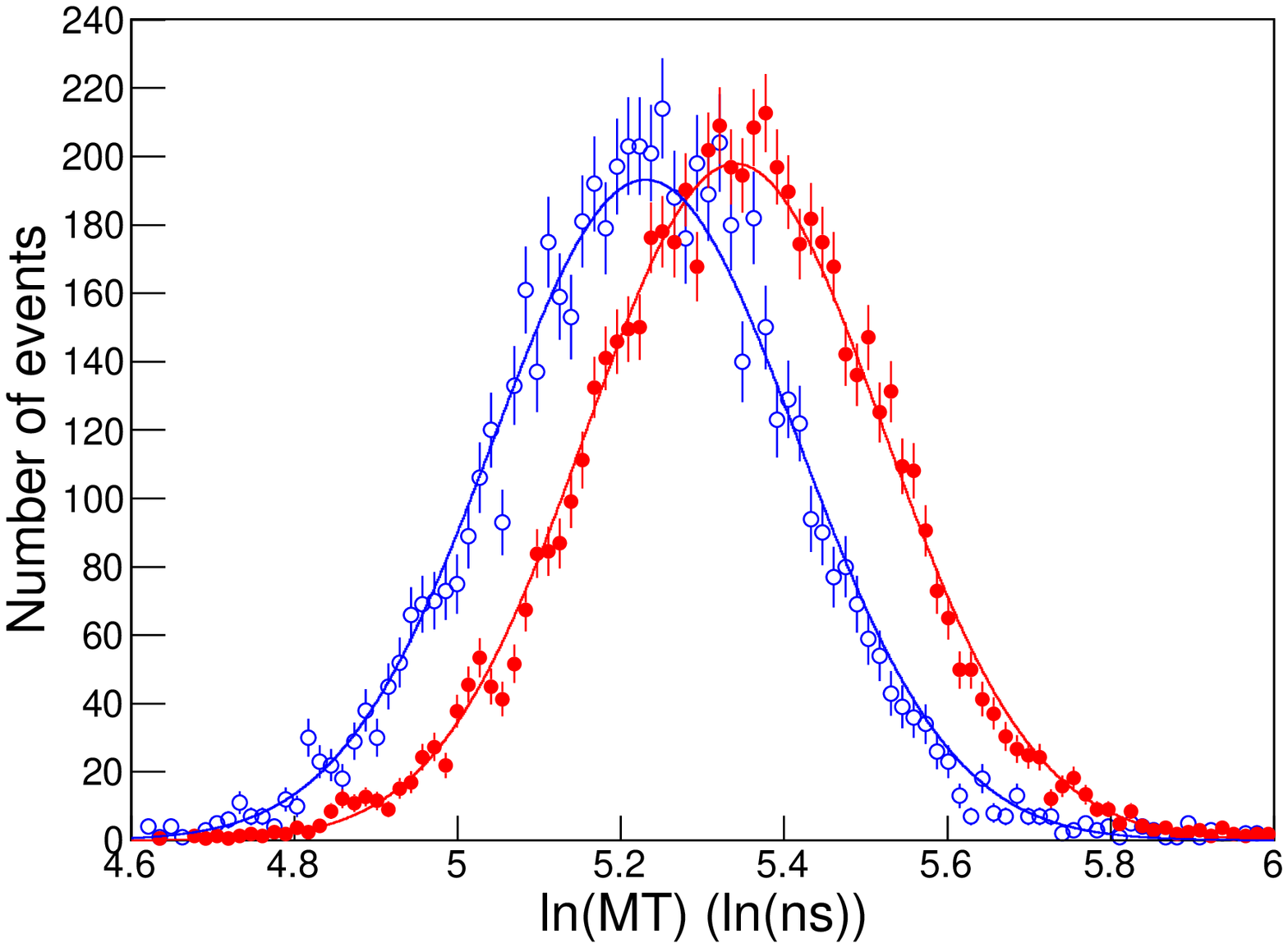}&
\includegraphics[width=0.45\textwidth]{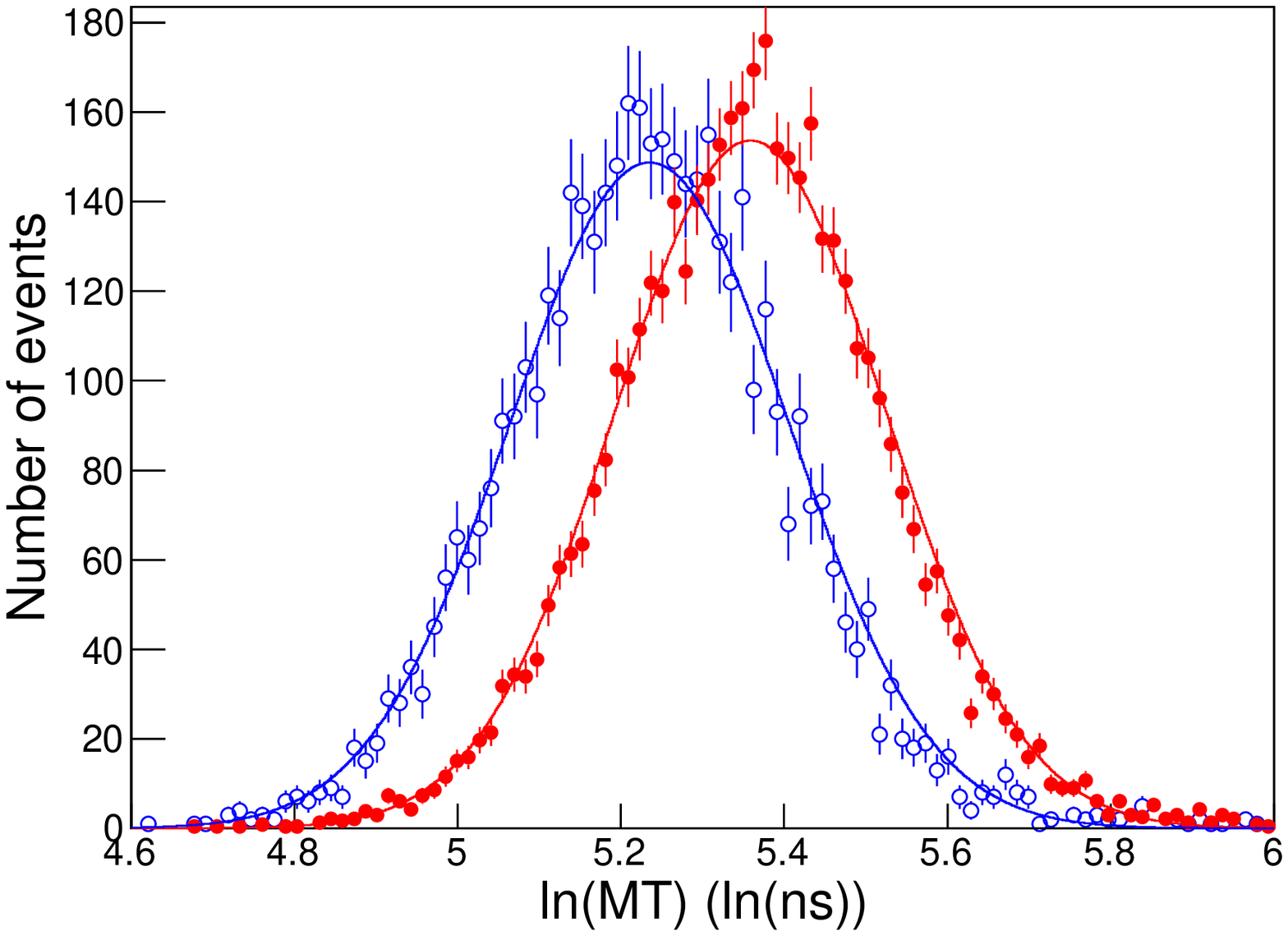}\\
(c) 3-4~keV& (d) 4-5~keV \\
\end{tabular}
\end{center}
\caption[Data fit]{
The ln(MT) distribution of neutron-induced events (open circle) from the Am-Be source and electron-induced events (filled circles) from the $^{137}$Cs source. The results of fits using asymmetric Gaussian functions are superimposed. 
}
\label{ref:psdkeV}
\end{figure*}

Because of the different time distributions of the photoelectrons produced by nuclear and electron recoils in scintillation crystals, it is possible to use PSD to distinguish the nuclear recoil signals from the electron-recoil-dominant backgrounds. To characterize the time distributions, we use the natural logarithm of the mean decay time~(ln(MT)), defined as 
\begin{equation}
\mathrm{ln(MT)}=\mathrm{ln}\left(\frac{\sum{A_it_i}}{\sum{A_i}}-t_0\right),
\end{equation}
where $A_i$ is the charge of the $i$th cluster, $t_i$ is the time of the $i$th cluster, and $t_0$ is the time of the first cluster. Here, only clusters within 1.5$\mu$s from $t_0$ are considered. Figure~\ref{ref:psdkeV} shows the ln(MT) distributions for the nuclear and electron recoil events in the NaI(Tl) crystal for 1~keV bins of visible energy for the range $1<E<5$~keV. In this figure, differences in the ln(MT) distributions between nuclear recoils from electron recoils are evident, even for energies as low as 1~keV. This result is significantly improved compared with those of previous studies that reported PSD capabilities only for energies of 4~keV and higher~\cite{gerbier,naiad,damapsd}. This gives us confidence to design an experiment for WIMP searches with NaI(Tl) crystals based on the PSD performance. 
 
\begin{figure*}
\begin{center}
\begin{tabular}{cc}
\includegraphics[width=0.45\textwidth]{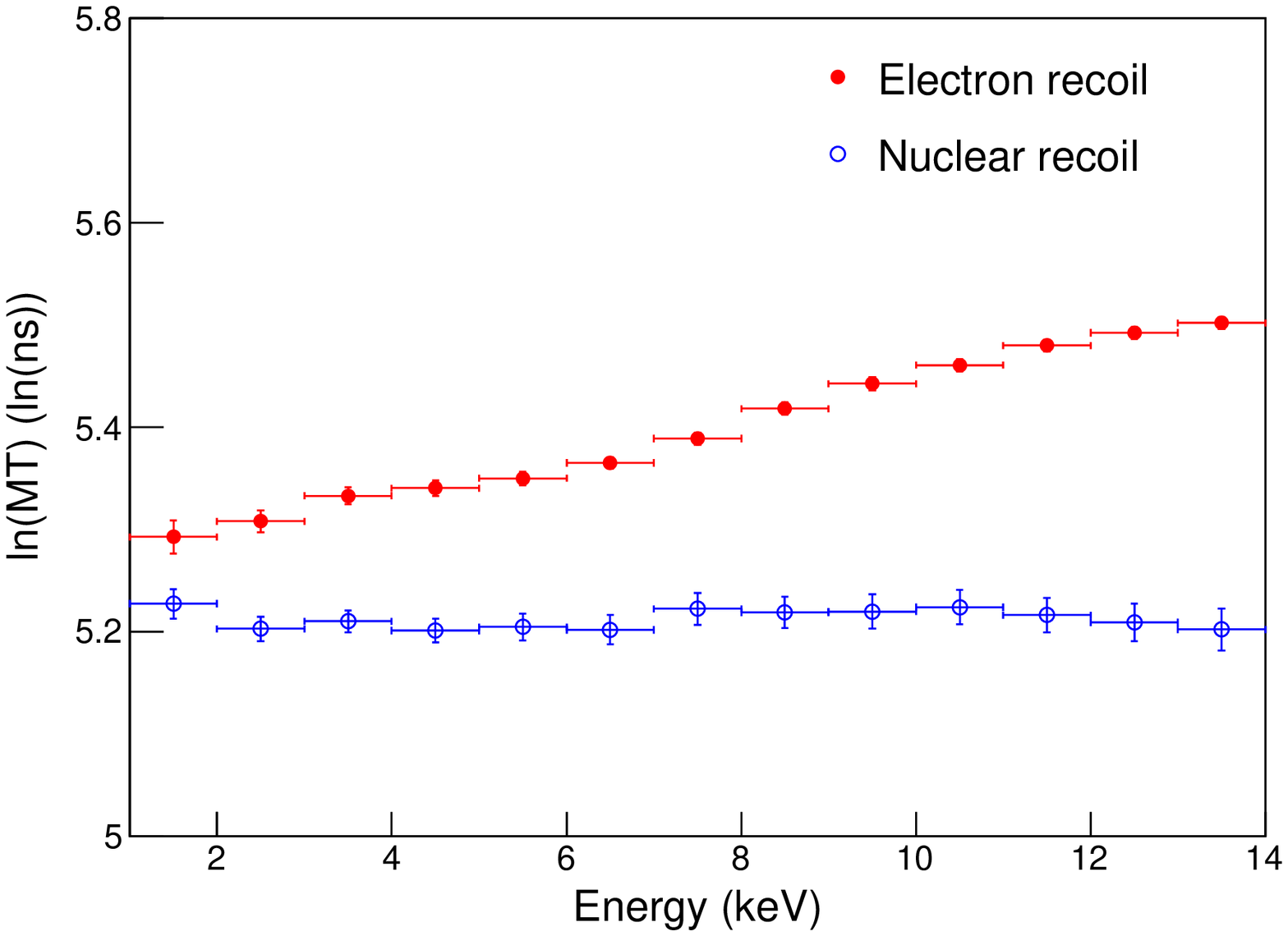}&
\includegraphics[width=0.45\textwidth]{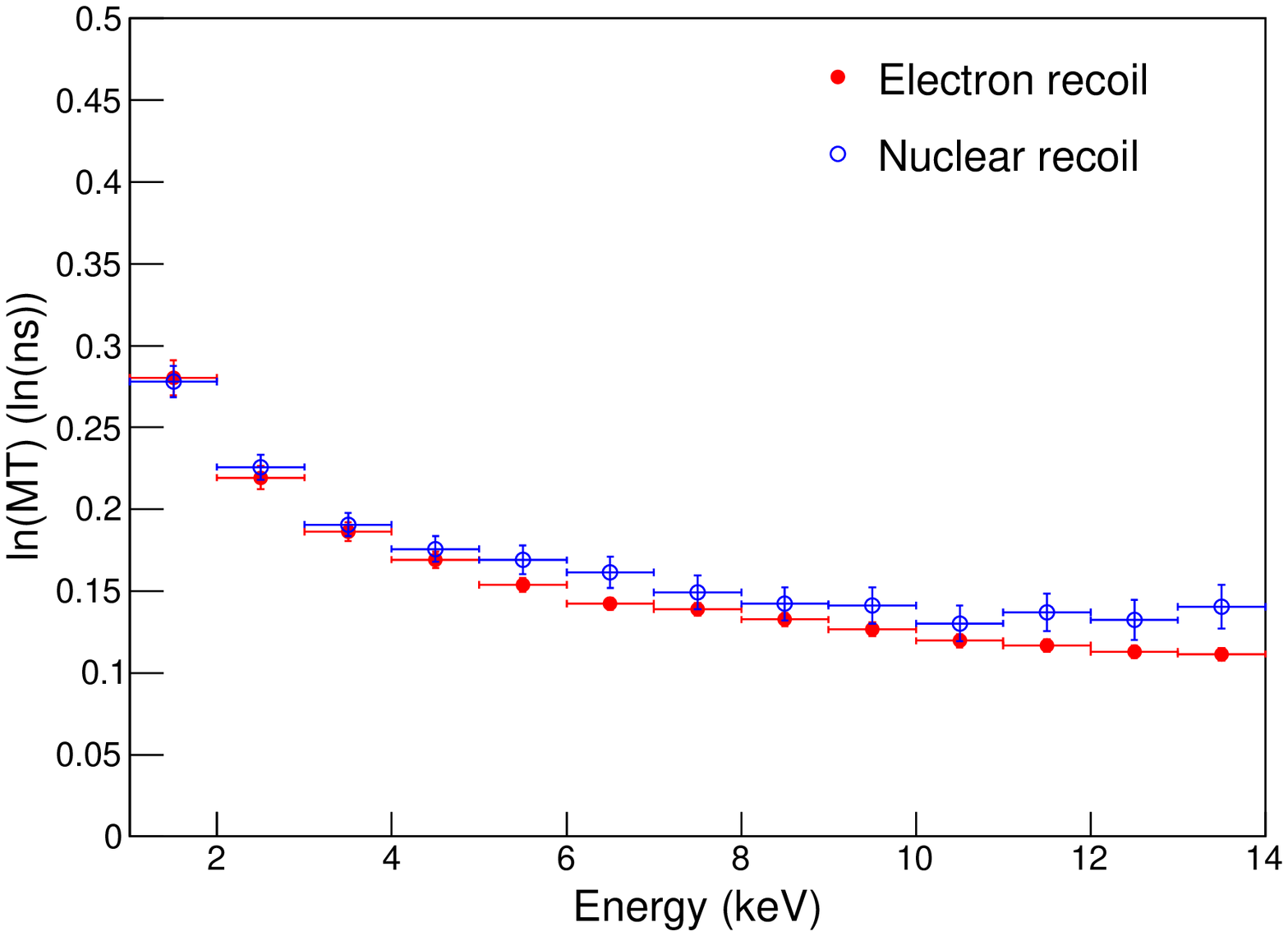}\\
(a)  & (b) \\
\end{tabular}
\caption[]{
(a) The most probable value of ln(MT) from the asymmetric Gaussian fits for each energy bin for neutron-induced (open circles) and electron-induced~(filled circles) events, respectively. 
(b) The root mean square~(RMS) of ln(MT) as a function of energy for neutron-induced~(open circle) and electron-induced~(filled circle) events. 
}
\label{ref:psd}
\end{center}
\end{figure*}

We performed fits to the ln(MT) distributions using asymmetric Gaussian functions that are shown in the plots, where good agreement with the data is observed. 
Similar fits were performed for each 1~keV energy bin from 1 to 14~keV for nuclear and electron recoil events, and Fig.~\ref{ref:psd} shows the most probable value and the root mean square~(RMS) of ln(MT) for electron and nuclear recoil data. As shown in the plots, the separation power is reduced at low energy, but some PSD capability persists at energies as low as 1~keV, which is even better than CsI(Tl) crystals~\cite{csi-psd1,csi-psd2}. 

\begin{figure}
\begin{center}
\includegraphics[width=0.8\columnwidth]{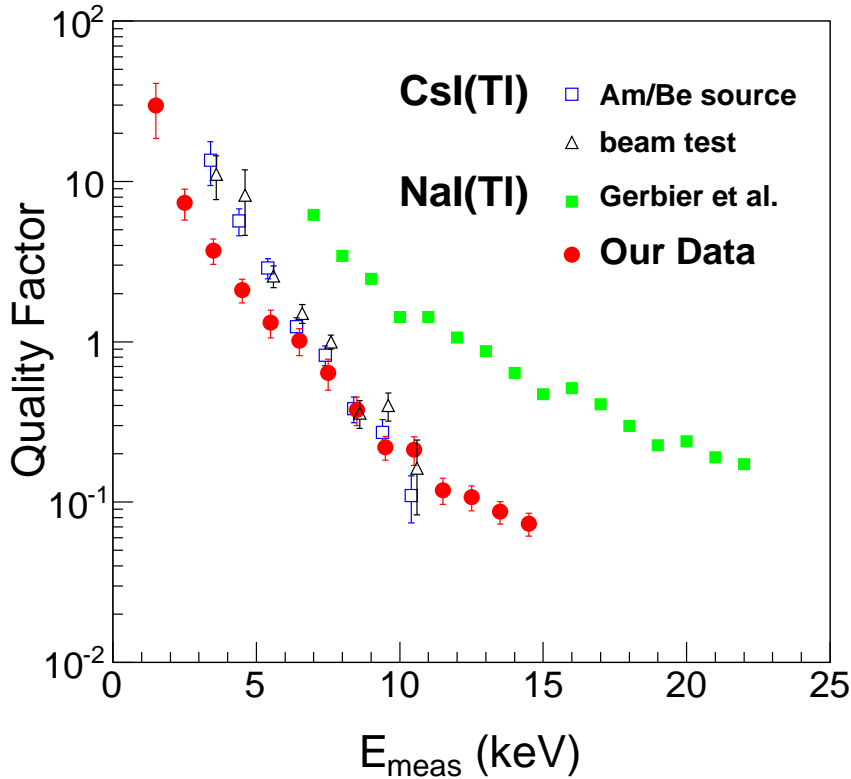}
\caption[]{The measured quality factor of the new NaI(Tl) crystal 
	compared to previous NaI(Tl) measurement by Gerbier {\it et al.}~\cite{gerbier} and CsI(Tl) measurements using monoenergetic neutron beam~\cite{csi-psd1} as well as the Am-Be source~\cite{csi-psd2}.
}
\label{ref:quality}
\end{center}
\end{figure}

We characterize the PSD power using a quality factor~\cite{quality} that is commonly used to characterize WIMP-search detectors~\cite{gerbier,csi-psd1,csi-psd2} and defined as
\begin{equation}
K \equiv \frac{\beta(1-\beta)}{(\alpha-\beta)^2},
\label{eq:quality}
\end{equation}
where $\alpha$ and $\beta$ are fractions of the nuclear and electron recoil events that satisfy the selection criteria, respectively,  where we have varied these criteria to provide the best~({\it i.e.}, the minimum) value of the quality factor.  For a detector with perfect discrimination, $\alpha=1$ and $\beta=0$, thus a smaller quality factor means a better PSD power.
Figure~\ref{ref:quality} shows the measured quality factor of the NaI(Tl) crystal studied here compared with results from previous NaI(Tl)~\cite{gerbier} and CsI(Tl) crystal measurements~\cite{csi-psd1,csi-psd2}. As shown in this figure, the quality factor of this new, high light-yield NaI(Tl) crystal is approximately one order of magnitude lower than that from previous studies~\cite{gerbier}. To demonstrate that this is mainly due to the approximately two times higher light yield of the new crystal, we adjust our results by assuming a light yield similar to typical NaI(Tl) crystals and find results that agree with previous measurements. We also find that the quality factor of this crystal is slightly better than that of CsI(Tl) crystals, especially at energies below 6~keV. This promises relatively good sensitivity for low-mass WIMP detection with these NaI(Tl) crystals.

\section{Expected sensitivity of a NaI(Tl) experiment using PSD}

The Korea Invisible Mass Search~(KIMS) Collaboration recently joined a consortium that is developing low-background, high light-yield NaI(Tl) crystals for attempts to reproduce the DAMA/LIBRA observation. We have achieved a background level of approximately 3~dru at 6~keV with a below 2~keV energy threshold~\cite{kims-nai}. Because we already identified the significant backgrounds of the NaI(Tl) crystals as being due to $^{210}$Pb and $^{40}$K contaminations, the background levels of future crystals that are produced as part of this program will be significantly reduced. Even without internal background reduction, it should be possible to achieve a $\sim$2~dru  background level by immersing the NaI(Tl) crystal array inside a liquid scintillator box that serves as an active veto. After growing NaI crystals with a total mass of 100~kg, which may be completed by the end of 2015, the KIMS collaboration plans to start the KIMS-NaI experiment that, after three years of stable operation, could unambiguously confirm or refute the DAMA/LIBRA annual modulation signals without any model dependence. 

In the meantime, we plan to extract information about the WIMP interaction using NaI(Tl) crystals as early as possible. One of our goals is to extract the WIMP interaction signal using data from the initial period of one year or less. 
Because of the very good PSD power of the new NaI(Tl) crystals, 
the extraction of a meaningful cross section limit for WIMP-nucleon interactions in NaI should be possible with a small data set. 
To quantify the performance of this measurement, we conservatively estimate the  sensitivity of the KIMS-NaI experiment by assuming a 100~kg NaI(Tl) crystal array with a 2~dru background, a 1-year exposure, and a 2~keV energy threshold. 

\begin{figure*}
\begin{center}
\begin{tabular}{cc}
\includegraphics[width=0.45\textwidth]{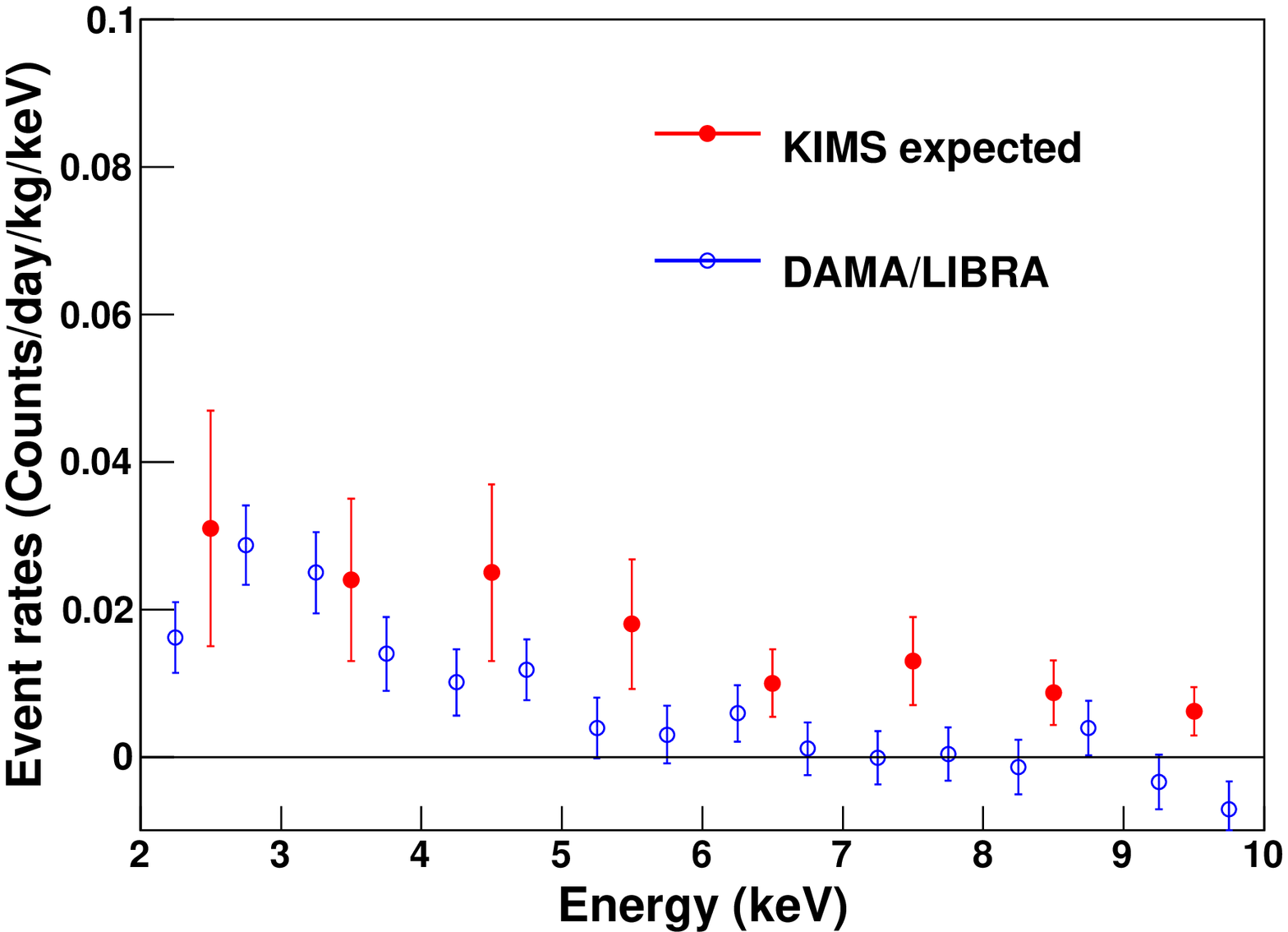}&
\includegraphics[width=0.45\textwidth]{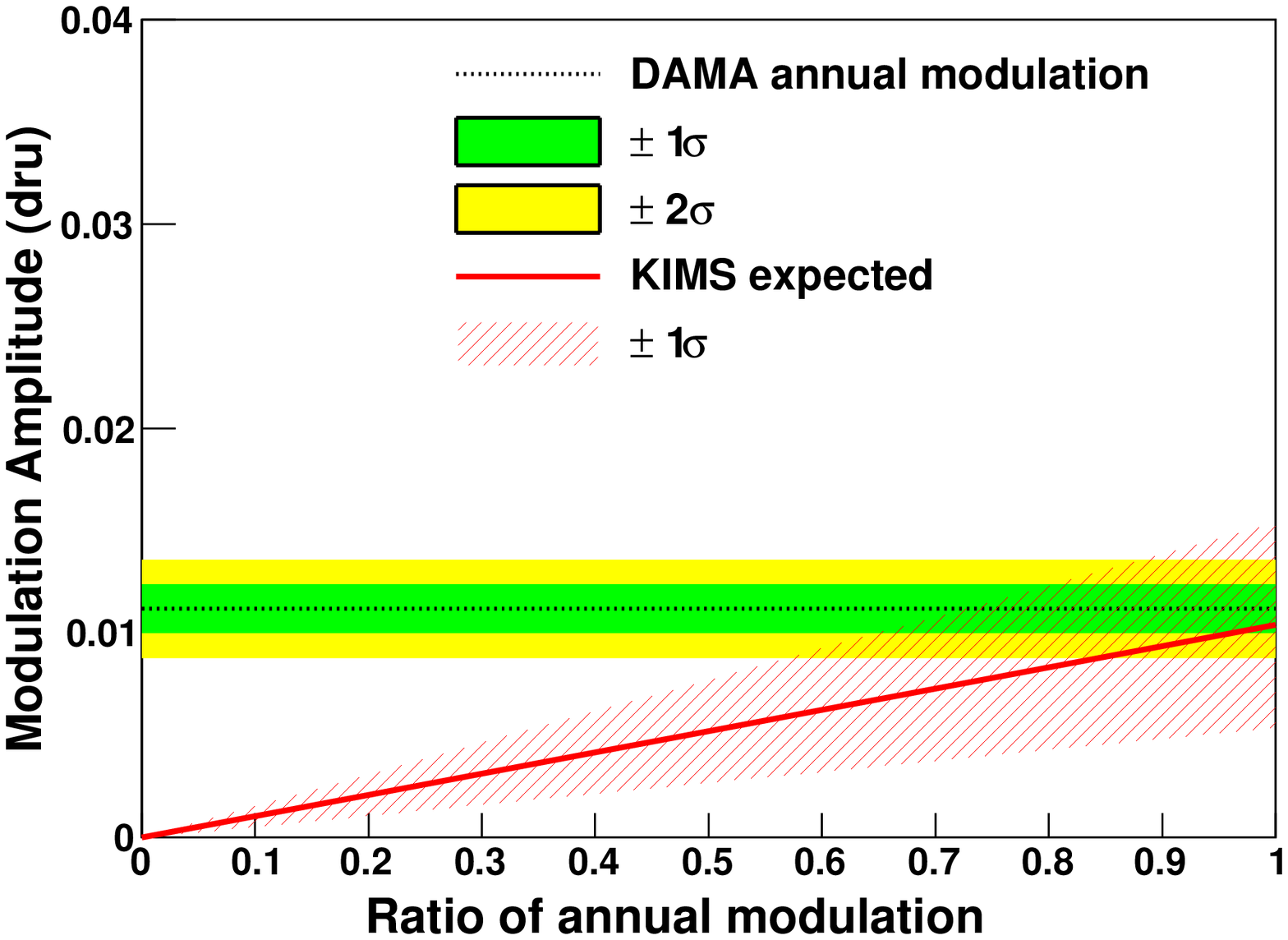}\\
(a)  & (b) \\
\end{tabular}
\caption[]{ (a) The median expected 90\%-confidence-level upper limits~(points) with 1$\sigma$ standard deviation probabilities~(error bars) on the rate of nuclear recoil events  
from the KIMS-NaI 1-year experiment~(filled circle) are compared with the annual modulation amplitudes observed by the DAMA/LIBRA
experiment~(open circle).  (b) Expected 90\%-confidence-level upper limit on the annual modulation amplitude as a function of the annual modulation ratio of WIMP-nucleon interaction from the KIMS-NaI experiment, which is compared with the DAMA/LIBRA observation in the 2-6~keV energy range.
}
\label{ref:keVlimit}
\end{center}
\end{figure*}

We performed simulated experiments to extract the upper limits of the nuclear recoil events. For events in each 1~keV bin, we ran 2,000 simulated experiments based on the expected electron recoil background, assuming no nuclear recoil signal events. We construct Bayesian likelihoods that are  formed as products of the measured log(MT) probability distributions for nuclear recoil signals and electron recoil backgrounds. For each simulated experiment, 90\% confidence level~(CL) limits are determined such that 90\% of the posterior densities of the nuclear recoil event rates fall below the limit. In Fig.~\ref{ref:keVlimit}~(a), we present the median expected 90\% CL upper limits~(points) together with 1$\sigma$ standard deviation probability~(error bars) of the nuclear recoil events for each 1~keV bin. The DAMA/LIBRA annual modulation amplitudes for each 0.5~keV bin from Ref.~\cite{savage} are overlaid in the plot for comparison. At low energies, the KIMS-NaI expected upper limits of the nuclear recoil rates corresponding to the WIMP interaction with NaI(Tl) nuclei are consistent with the annual modulation amplitudes of the DAMA/LIBRA experiment. 

To make a direct comparison between the expected nuclear recoil rate from the KIMS-NaI experiment and the annual modulation amplitudes observed by the DAMA/LIBRA experiment, we need to know the expected ratio of the annual modulation amplitude to the total WIMP-nucleon interaction rate, which depends on the models of the halo structure as well as the underlying theory of the WIMP candidate. In other words, a WIMP-nucleon interaction limit determined using the PSD analysis, can constrain the ratio of the annual modulation amplitude to the total rate of the WIMP-nucleon interaction, which is model-independent. The annual modulation amplitude observed by the DAMA/LIBRA experiment for the 2--6~keV events with the highest significance is 0.0112$\pm$0.0012~dru~\cite{DAMA3}. The expected 90\% CL upper limit of the nuclear recoil rate from the KIMS-NaI experiment is 0.0104$\pm$0.0050~dru. In Fig.~\ref{ref:keVlimit}~(b), we convert the expected upper limits of the nuclear recoil rate from the KIMS-NaI experiment into an annual modulation amplitude as a function of the annual modulation ratio from the total WIMP-nucleon interaction rate. We find that the KIMS-NaI experiment can investigate the allowed parameter space of the DAMA 2$\sigma$ signal region for any model with an annual modulation ratio less than 85\%, assuming the median expectation shown in Fig.~\ref{ref:keVlimit}~(b). 

\begin{figure}
\begin{center}
\includegraphics[width=0.8\columnwidth]{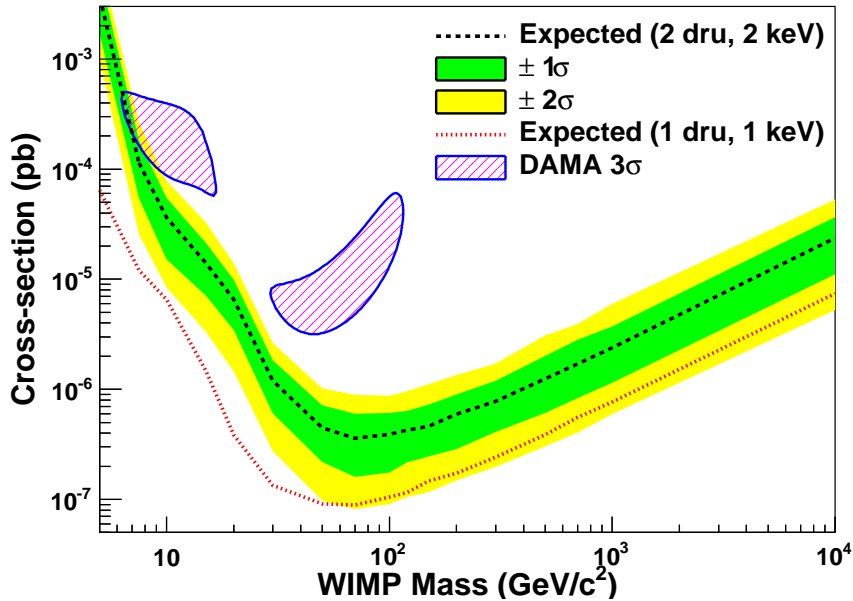}
\caption[]{
The median expected~(black dotted line) 90\%-confidence-level upper limit on 
the WIMP-nucleon spin-independent cross-section assuming the background-only hypothesis,
shown together with WIMP-induced DAMA/LIBRA 3$\sigma$-allowed region~(solid contour). The dark~(green) and light~(yellow) shaded bands indicate
the 1$\sigma$ and 2$\sigma$ standard deviation probability regions in which the limits
are fluctuated. The red dotted line indicates the expected limit
assuming an optimistic background level of 1~dru and an energy threshold of 1~keV.
}
\label{ref:limit}
\end{center}
\end{figure}

Assuming the standard halo model~\cite{shm}, we estimate the sensitivity of the KIMS-NaI experiment as the predicted cross-section limits for the WIMP-nucleon spin-independent interactions in case of no signal for an ensemble of Monte-Carlo experiments. For each experiment, we determine simulated log(MT) distribution for an electron-recoil-background only hypothesis. The rate of background, assuming a 1-year exposure of a 100~kg array of NaI(Tl) detectors with 2~dru background rate, has Poissonian fluctuations. 
We then fit the log(MT) spectrum with signal plus background hypothesis with flat priors for both the signal and the background rate.
We construct a Bayesian likelihood for each 1~keV bin from fits to the log(MT) distribution. The overall likelihood is obtained by multiplying the likelihood for each bin between 2 and 10~keV considering the energy-dependent rate of WIMP interaction for WIMP masses between \gevcc{5} and \gevcc{10000}. 
The 90\% CL limit for each simulated experiment is determined such that 90\% of the posterior density of the WIMP-nucleon cross-section falls below the limit. 
The median expected 90\% CL limits and 1$\sigma$ and 2$\sigma$ standard-deviation probability regions are calculated from the result of 2,000 simulated experiments. 
To directly compare these results with the DAMA/LIBRA allowed signal region, we use quenching factors from Ref.~\cite{savage} of 0.3 and 0.09 for Na and I, respectively. 
The quenching factor is the fraction of the recoil energy deposited by a WIMP with respect to the energy deposited by an electron~(or $\gamma$-ray).  
Figure~\ref{ref:limit} shows the expected median limit, and 1$\sigma$ and 2$\sigma$ standard deviation probability regions, of the KIMS-NaI experiment using PSD analysis. 
We also include results from an optimistic KIMS-NaI expected limit of a 1~dru background rate with a 1~keV energy threshold. 
These results are further compared with the DAMA/LIBRA 3$\sigma$ allowed signal region from Ref.~\cite{savage}. 
As shown in this figure, the median expected limit from the KIMS-NaI 100~kg$\cdot$year data can investigate the DAMA/LIBRA 3$\sigma$ regions even for the conservative threshold and background level assumptions.
This comparison, based on measurements using the same target-detector material, may provide a better understanding of the nature of dark matter and of the DAMA/LIBRA signature. 
If the KIMS-NaI experiment obtains similar annual modulation signatures using NaI(Tl) crystals in the future, this PSD measurement will provide valuable information about the annual modulation events that will help determine whether or not they originate from WIMP-induced nuclear recoil events.  

\section{Conclusion}
We report measurements of the PSD characteristics for  the nuclear and electron recoil events of the NaI(Tl) crystal. Because of the high light output of recently developed NaI(Tl) crystals, we find a PSD quality factor that is approximately one order of magnitude better than those of  previously studied crystals. The influence of this good PSD capability on the expected sensitivity of the KIMS-NaI experiment is evaluated. Model independent and dependent comparisons of the DAMA/LIBRA annual modulation signature give confidence that the KIMS-NaI experiment can investigate interesting regions of parameter space with one year of accumulated data.
 
\begin{acknowledgments}
This research was funded by the Institute for Basic Science~(Korea) under project code IBS-R016-D1 and was supported by the Basic Science Research Program through the National Research Foundation of Korea funded by the Ministry of Education (NRF-2011-35B-C00007). 
\end{acknowledgments}

\end{document}